\documentclass[12pt,a4paper,final]{iopart}

\usepackage{iopams}
\usepackage{graphicx}
\usepackage{cite}
\begin{document}

\title[Study of extreme states of matter at high energy densities]
{Study of extreme states of matter at high energy densities and high strain rates with powerful lasers}

\author{I~K~Krasyuk$^1$, P~P~Pashinin$^1$, A~Yu~Semenov$^1$, K~V~Khishchenko$^2$ and V~E~Fortov$^2$}

\address{$^1$~A~M~Prokhorov General Physics Institute RAS, Vavilova 38, Moscow 119991, Russia\\
$^2$~Joint Institute for High Temperatures RAS, Izhorskaya~13 Bldg~2, Moscow 125412, Russia
}
\ead{krasyuk99@rambler.ru}
\vspace{10pt}
\begin{indented}
\item[]May 2016
\end{indented}

\begin{abstract}
In the paper, a review of most important results of experimental studies of thermonuclear plasma in conical targets, generation of shock waves and spallation phenomena in different materials, which were carried out at laser facilities of the A~M~Prokhorov General Physics Institute RAS since 1977, is presented.
\end{abstract}

%
%
%
%
%

\section{Introduction}
Since 1977, at the A~M~Prokhorov General Physics Institute (GPI) RAS, in collaboration with other institutes of the Russian Academy of Sciences (Joint Institute for High Temperatures (JIHT) RAS, Institute of Problems of Chemical Physics (IPCP) RAS, A~A~Dorodnitsyn Computing Center RAS, L~D~Landau Institute of Theoretical Physics RAS) as well as the Moscow Institute of Physics and Technology, investigations of properties of matter under extreme conditions (high temperatures, high pressures, high strain rates) are carried out~\cite{1, 1a, 4, Bespalov-1984, Bushman-1984, Bushman-1986, 5, 10, 2, 2a, 9, Krasyuk-1997, 3, 11, 3a, 11a, 11b, 8h, 11c, 6, Vovchenko-QE-2007, Vovchenko-PF-2009, 15, 12, 12a, 13, 13a, 14, 14a, 14b}. To this end, unique laser facilities based on the neodymium glass ``Phoenix'', ``Sirius'', ``Kamerton'' and ``Kamerton-T'' were set up. The Nobel Prize winner academician A~M~Prokhorov was interested in and supported continuously these works, and it contributed successful implementation of the tasks. In the paper, most important results obtained during the investigations are summarized.

\section{Generation of thermonuclear plasma in conical targets}
\subsection{Laser action on conical targets}
Experiments on generating thermonuclear plasma in conical targets have been carried out at laser facility ``Phoenix'' based on neodymium glass (the wavelength $\lambda=1.06$~$\mu$m, the pulse duration $\tau=20$~ns, the energy up to 100~J per pulse, the laser irradiation intensity up to $10^{11}$~W/cm$^2$). At once, encouraging result was obtained: neutron yield of DD reaction at every experiments exceeded $10^4$~particles per pulse~\cite{1, 1a}. A photograph of a target microsection after the laser action is shown in figure~1. Spherical cavity in the vertex of cone appeared as a result of expansion of gas compressed at pressures up to 2~TPa when the generated shock wave had been converged at the vertex (a zone of cumulative explosion is shown schematically at the cavity center). During these investigations, measured were the following values: the velocity of the target envelope, the part of laser energy deposited into the target, the rate of evaporation of the envelope, the ablation pressure acting on the envelope, the neutron yield as a function of initial conditions over a wide range of its variation.

\begin{figure}[b]
\centering\includegraphics[width=0.67\columnwidth]{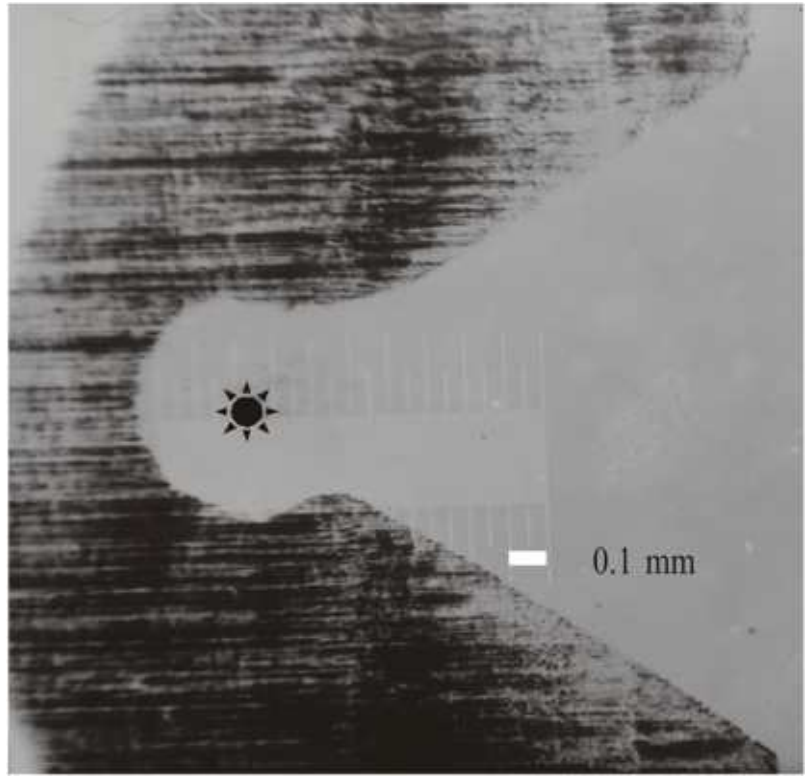}
\caption{The conical target microsection after the action by laser pulse of 60~J on the envelope of polyethylene terephthalate with thickness 5~$\mu$m. Input diameter of the target is 1.7~mm. Ablation pressure is 2.4~GPa. Before the shot, the target was filled in by gaseous deuterium at the pressure of 0.05~MPa. Measured neutron yield is $2.3\times 10^4$.}
\end{figure}

Large number of performed experiments enabled to carry out a statistical treatment and make important conclusions about the process of heating of thermonuclear plasma into the target \cite{2, 2a}. It was ascertained that the amount of thermonuclear fuel heated up to temperatures of 10~keV comes to 0.15\% of total mass of initial filling of the target. It was supposed that, at conical targets, cumulative phenomena at central part of the cone plays crucial role in plasma heating up to thermonuclear temperatures. Subsequent theoretical investigations using two-dimensional numerical simulations~\cite{3, 3a} had validated this supposition. As the simulations show, the main neutron yield happens during the stage of converging at the vertex of the cone cavity of the shock wave, which is generated in deuterium at the target envelope movement (figure~2).

\begin{figure}[t]
\centering\includegraphics[width=0.85\columnwidth]{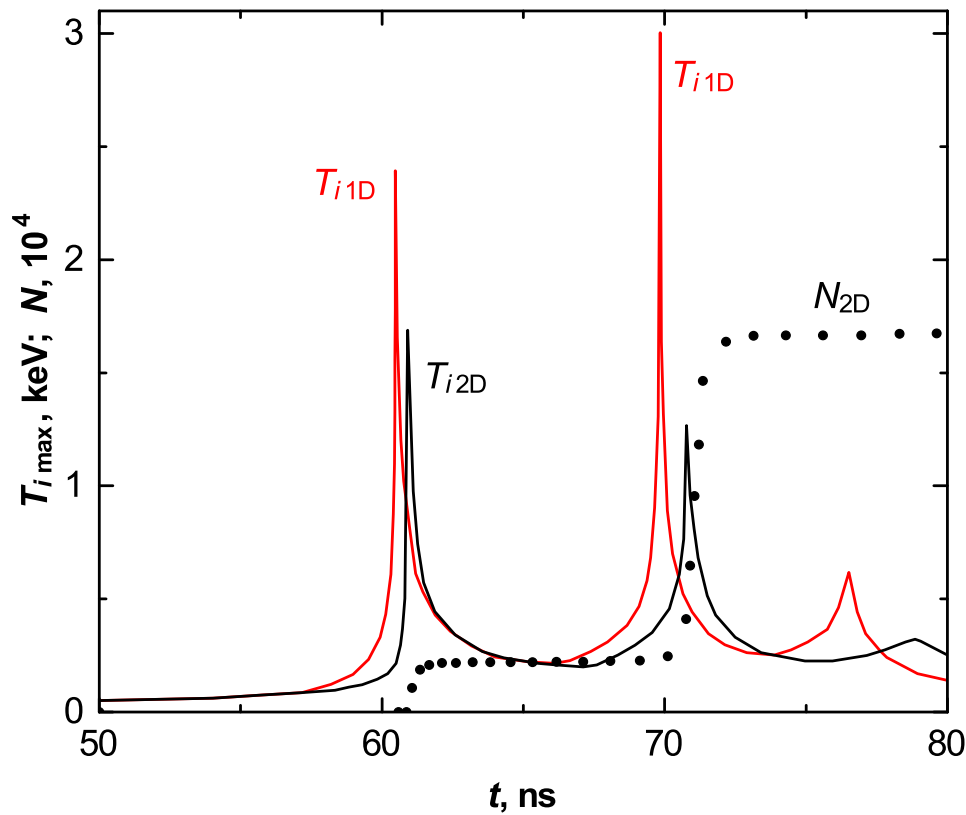}
\caption{Simulated results for experiment on laser compression of deuterium at the conical target: solid lines correspond to the maximum ion temperature from one- ($T_{i\,\mathrm{1D}}$) and two-dimensional calculations ($T_{i\,\mathrm{2D}}$); circles show neutron yield from two-dimensional calculation ($N_\mathrm{2D}$).}
\end{figure}

\subsection{Explosive action on conical targets}
After successful experiments with laser action on the conical targets, an idea was suggested to use such targets for quasispherical shock compression of thermonuclear plasma. New experiments were carried out by collaborators from GPI RAS and IPCP RAS \cite{4}. At the experiments, traditional explosive planar-projectile systems were used up to aluminum-flyer-plate velocities of 5.4~km/s with a simple scheme and of 18~km/s with use of a layer system. A stable yield was measured up to $10^7$ neutrons per shot. For understanding of possible mechanisms of deuterium-gas heating, both analytical calculations and numerical simulations were performed for the processes under study. The problem is sufficiently complex; it does not solved in full measure up to now. Analytical evaluation gives for maximum values of pressure of deuterium in the vertex of cone 6~TPa, compression ratio---3400 (a ratio of density to its initial value), temperature---0.24~keV, neutron yield---$10^5$. Discrepancy of the neutron yield with experimental value is connected to all appearance with complex structure of real hydrodynamic flow in the cone, for example, with Mach waves or jet streams. A possibility of jet streams formation is confirmed by two-dimensional simulations of hydrodynamic processes in conical targets (without gas filling) \cite{5} at the flyer plate with velocity 5.4~km/s and two thicknesses (2 and 0.25~mm). In the both cases, plate is made of aluminum, the target---lead; the cone base diameter is 2~mm, the cone angle is $28^\circ$. Results of simulations are shown in figure~3.

\begin{figure}[t]
\centering{\includegraphics[width=0.83\columnwidth]{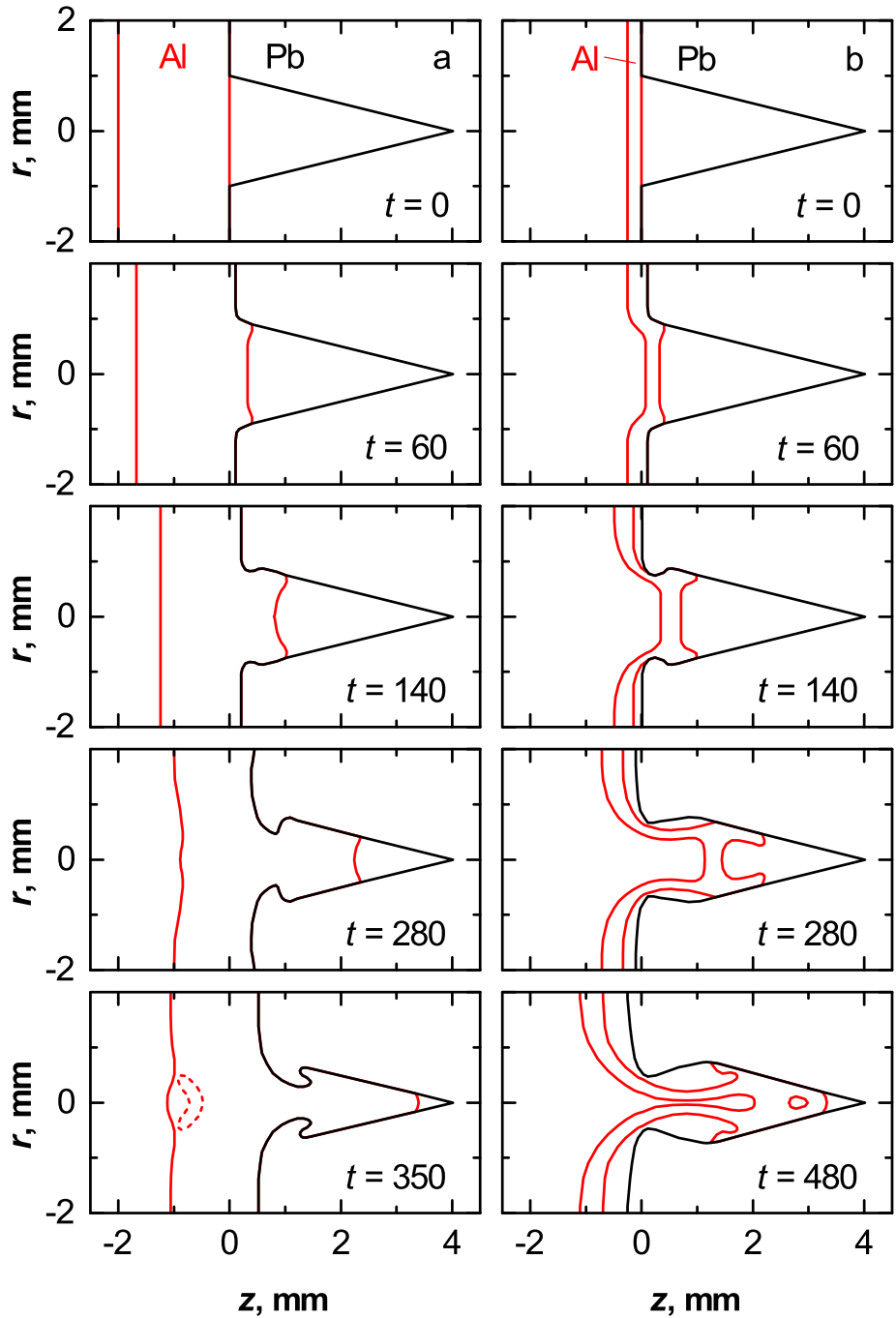}}
\caption{Cross-section of lead target with empty conical cavity under the action of aluminum flyer plate at different time moments ($t$, ns) from numerical simulations \cite{5}: flyer velocity 5.4~km/s, thickness (a) 2 and (b) 0.25~mm. Time is referenced from the moment $t=0$ when the flyer has arrived to the target surface. Dashed line indicates the spallation region.
}
\end{figure}

In the first case (thicker flyer plate), one can supposes that the gaseous deuterium placed in the target will be compressed and heated quasi-adiabatically. In the second case (thinner flyer plate), a cumulative effect becomes apparent. It seems that the junction of jet streams initiates the shock wave in the gas, the collapse of the wave leads to plasma formation with thermonuclear synthesis conditions. 

\subsection{Resuming remarks}

More details of using conical targets in investigations of inertial thermonuclear synthesis can be found elsewhere~\cite{Krasyuk-1997, 6}.

Recently, interest to conical targets is renewed in connection with an idea to use miniature conical targets for fast ignition of laser thermonuclear spherical targets~\cite{7}.

Using of conical targets for graphite-to-diamond transformation in a convergent shock wave was proposed~\cite{8} and investigated in numerical simulations~\cite{8a, 8b} and explosive experiments~\cite{8c}. As a result of the simulations, unexpected feature of convergent shock waves in porous media was discovered~\cite{8d, 8e, 8f}, which is revealed particularly in threefold increase of maximum pressure in the target at decrease of initial density of the graphite cone from 2.26 to 1.7~g/cm$^2$.

Conical targets are complex object for experimental study because the diagnostics of hydrodynamic processes within the targets is difficult. However, current development of numerical methods allows modeling of physical processes in conical targets in a wide range of initial conditions for the purpose of realizing of required characteristics of thermonuclear plasma in experimental conditions.

\section{Laser generation of strong shock waves}
One more direction of investigations at GPI RAS is using powerful laser pulses for generation of strong shock waves and their application to study of thermophysical and mechanical properties of matter under conditions of high pressures and high strain rates~\cite{8g, 8h}.
Note that fiber light guides elaborated due to the initiative by A M Prokhorov were used for the first time at diagnostics of shock waves generated by laser, explosive and electron-beam action~\cite{Bespalov-1984}. In the case of explosive experiments, using of the fiber light guides is of exceptional importance for safety of recording instruments.

\subsection{Dependence of ablation pressure upon the intensity of laser irradiation}
The problem of determining the ablation pressure on irradiated surface as a function of the intensity of laser irradiation was solved experimentally \cite{9}. This function is widely used in experiments on laser generation of shock waves. For the determining, different methods were applied such as measurement of time of shock-wave arrival on rear surface of the target, interference and Doppler registration of movement of foils in cylindrical and conical channels~\cite{10}. As a result, a dependence of the ablation pressure $P_{a}$ (TPa) upon the laser intensity $I_{l}$ (TW/cm$^2$) was proposed over the range from 0.04 to 1000~TW/cm$^2$:
\begin{equation}
P_{a}=\left\{\!\!
\begin{array}{ll}
1.2(10^{-2} I_{l})^{2/3}\lambda^{-2/3}[A/(2Z)]^{3/16} & \mathrm{at~}4.3 < I_{l} \leqslant 1000,\\
1.62(10^{-2} I_{l})^{7/9}\lambda^{-3/4} & \mathrm{at~}0.8 < I_{l} \leqslant 4.3,\\
1.7(10^{-1} I_{l})^{3/2}\lambda^{-3/4} & \mathrm{at~}0.04 \leqslant I_{l} \leqslant 0.8,
\end{array}
\right.
\end{equation}
where $\lambda$ is the laser irradiation wavelength ($\mu$m), $A$ and $Z$ are the atomic mass (u) and the atomic number of the target material respectively.

In connection with the problem of using of laser shock waves to obtain Hugoniot data for materials, the temperature behind the shock front was calculated \cite{11}. It was shown in particular that the preheating of the target by x-ray emission from ablation plasma may achieve 0.3~eV at laser intensities of 30~TW/cm$^2$. The preheating changes parameters of initial state of the target. It should be taken into account at analysis of results of experiments with laser generated shock waves.

\subsection{Using of laser for study of spallation phenomena in matter}
At GPI RAS, experimental study of spallation phenomena in different materials was started from investigation of possibilities of using of aluminum-magnesium alloy AMg6M for counter-meteor defense of the spacecrafts Vega-1 and Vega-2. At that time, for this purpose, the laser facilities ``Kamerton'' and ``Sirius'' were used. By now, with the laser ``Kamerton-T'', spallation phenomena are studied in aluminum, tantalum, copper, tungsten, palladium, lead, silicon, graphite, synthetic diamond and polymethylmethacrylate \cite{11a, 11b, 11c, 15, 12, 12a, 13, 13a, 14, 14a, 14b, Vovchenko-QE-2007, Vovchenko-PF-2009}. Photographs of rear sides of the targets of some materials after laser action are shown in figure~4.

\begin{figure}[t]
\centering{a~\includegraphics[width=0.45\columnwidth]{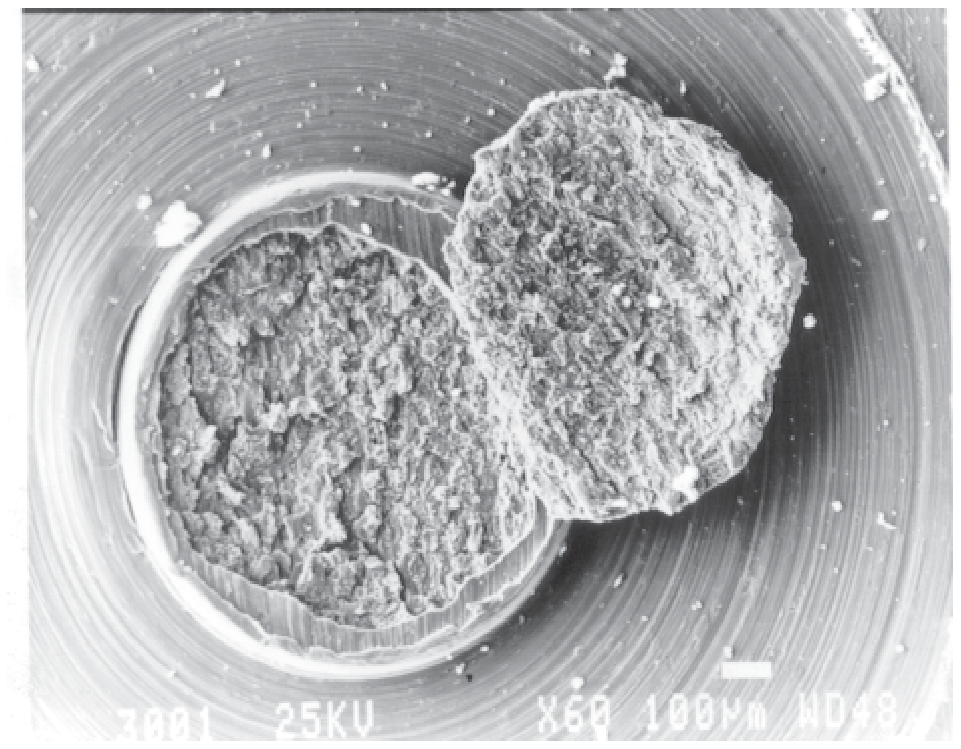}
b~\includegraphics[width=0.45\columnwidth]{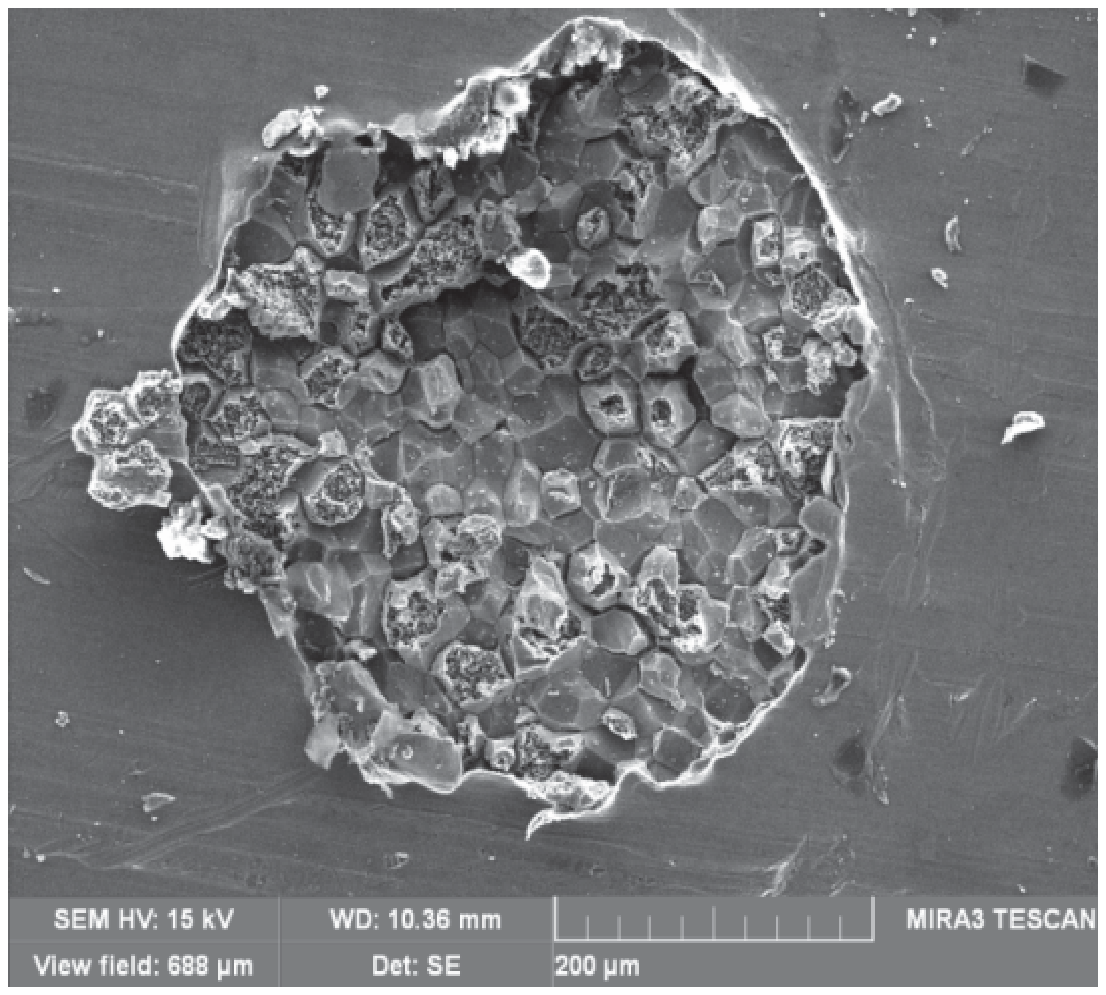}
c~\includegraphics[width=0.45\columnwidth]{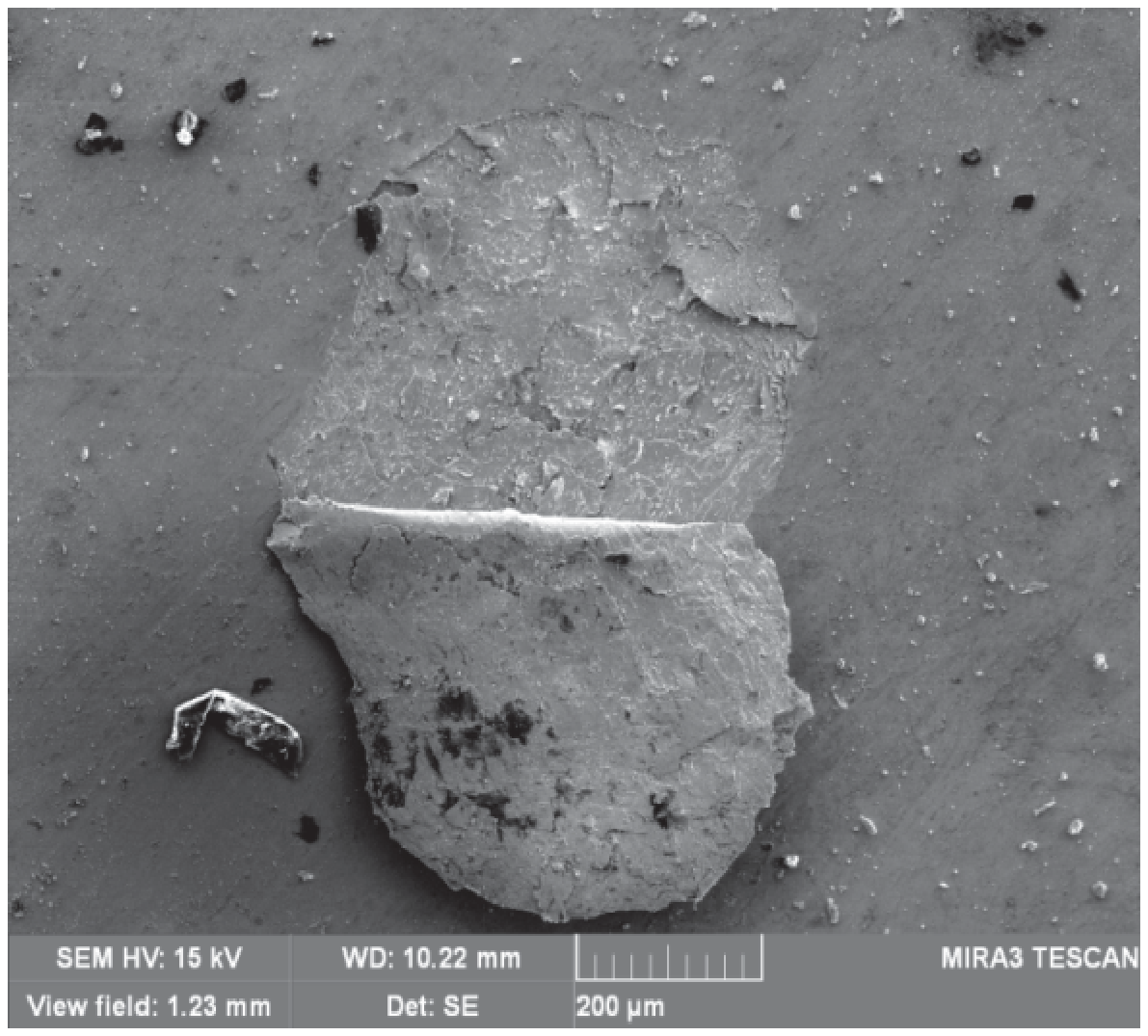}
d~\includegraphics[width=0.45\columnwidth]{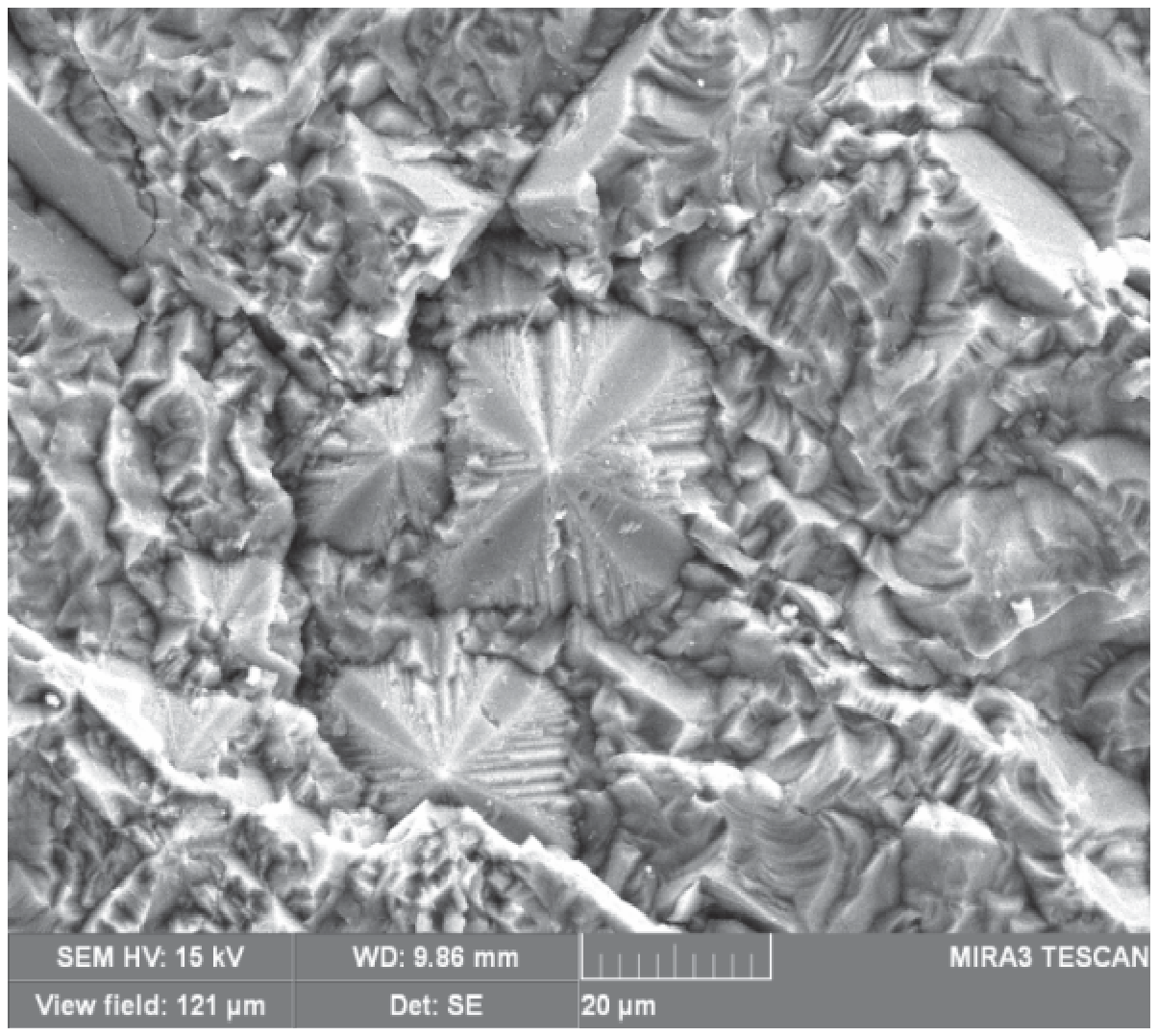}
e~\includegraphics[width=0.45\columnwidth]{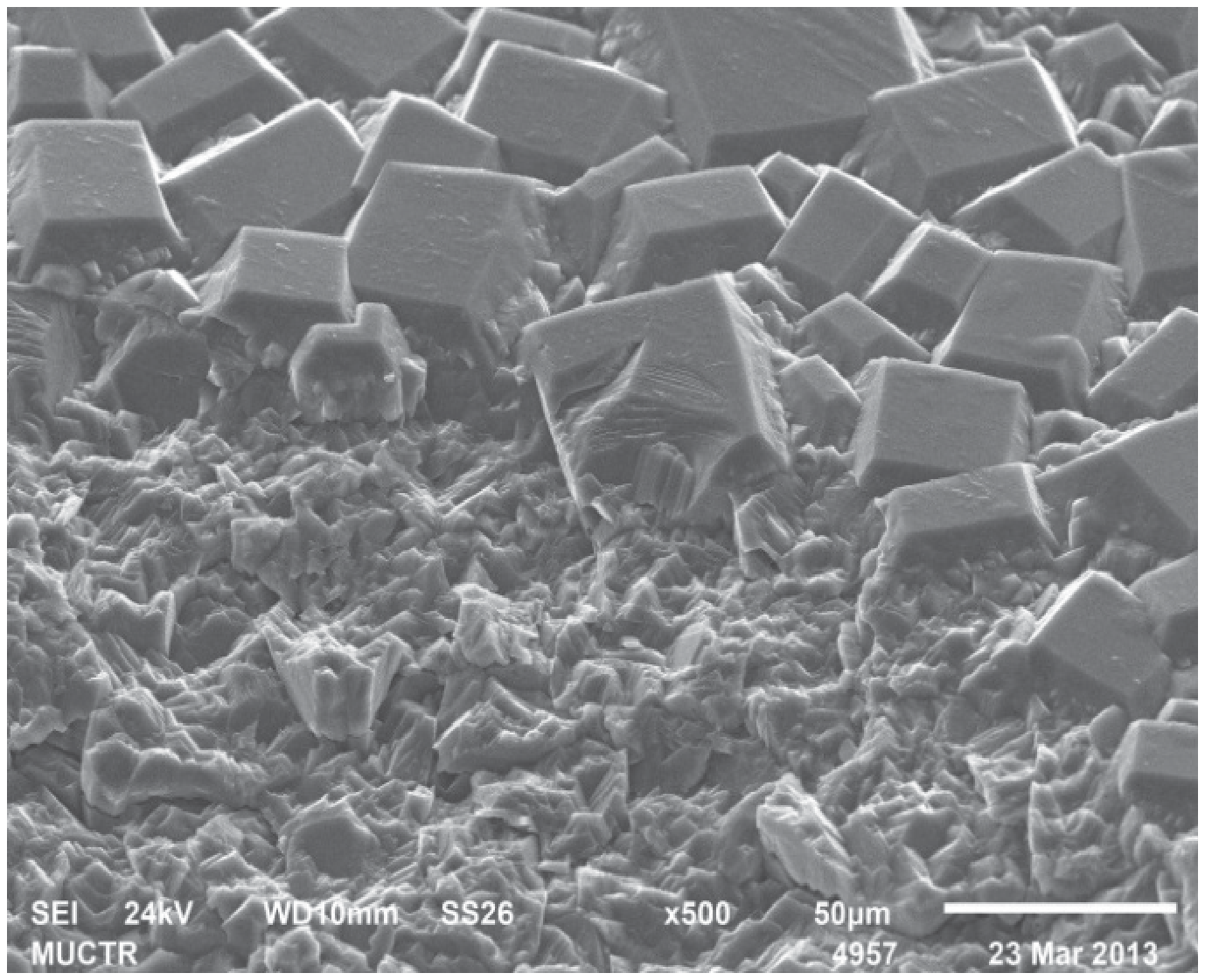}
f~\includegraphics[width=0.45\columnwidth]{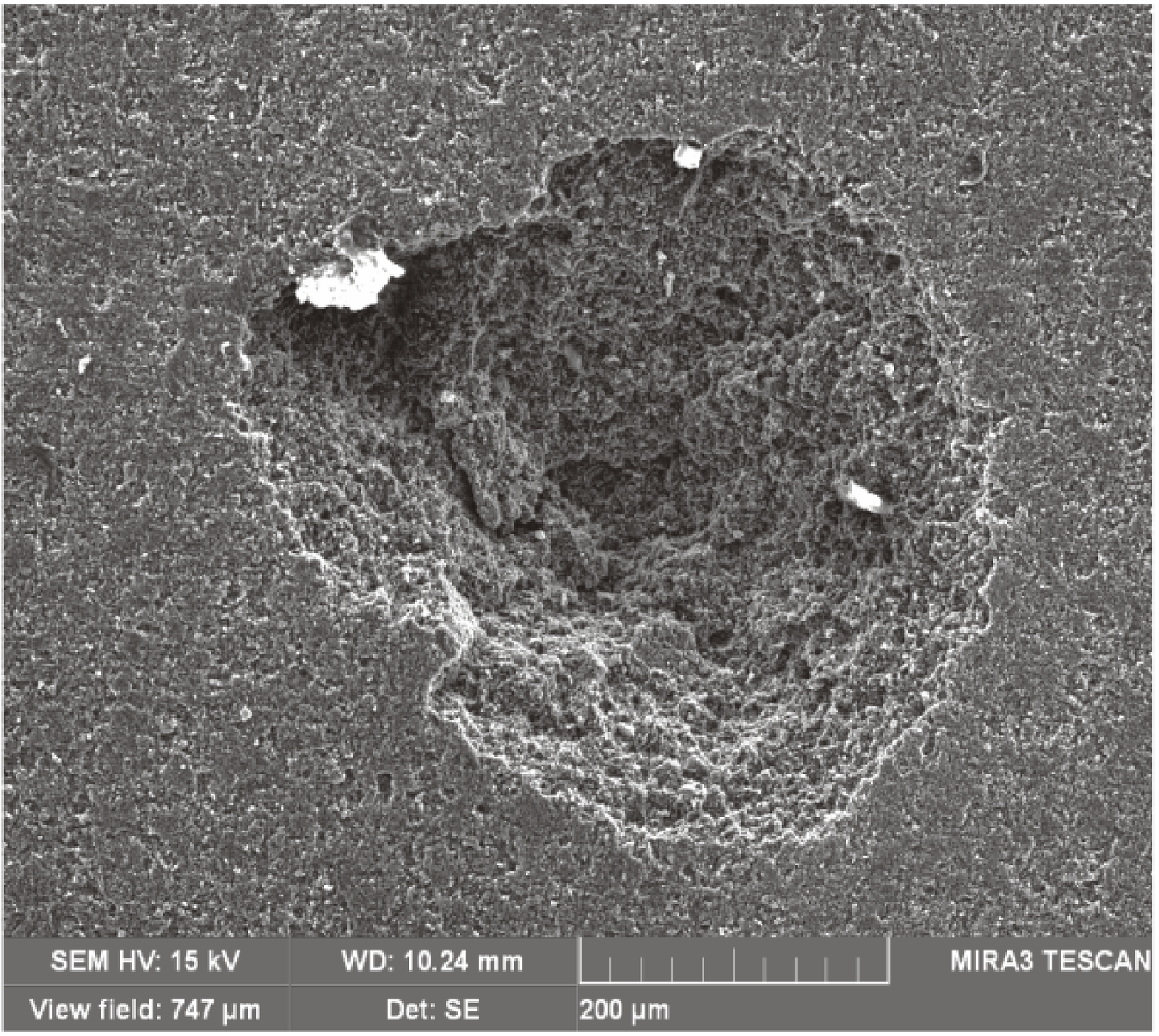}}
\caption{Photographs of rear sides of the targets obtained using a scanning electron microscope: a---aluminum-magnesium alloy AMg6M, b---aluminum, c---tungsten, d---silicon, e---synthetic diamond, f---graphite.}
\end{figure}

\subsection{Method of determining spall strength and strain rate at negative pressures}
To determine spall (tensile) strength $\sigma_\mathrm{sp}$ of a material at some strain rate $\dot{V}/V_0=\mathrm{d}(V/V_0)/\mathrm{d}t$ ($V_0$ is the initial specific volume, $\dot{V}$ is the rate of its changing with time $t$), an approach has been used based on measuring the depth of spall cavity $h$ after laser pulse action and subsequent modeling of shock-wave process in sample under study. For calculating of values $\sigma_\mathrm{sp}$ and $\dot{V}/V_0$, a one-dimensional computational code was developed and used, which is based on Courant--Isaacson--Rees scheme for hydrodynamic equations~\cite{16, 16a}. The code is supplemented with wide-range equations of state of materials in question~\cite{16b, 17, 17a, 17b, 17c, 17d}. In calculations, it was suggested that the time profile of the ablation pressure pulse on the irradiated surface of the target replicates the profile of laser pulse intensity. Relation between the amplitude of ablation pressure $P_{a}$ and the maximum intensity of laser irradiation $I_{l}$ is determined by equation~(1).

At the experiments, a value of the laser pulse intensity is registered, at which the spallation takes place. Then, the numerical simulation of shock wave propagation though the target is carried out. Results of a variant of the simulation are shown in figure~5. The moment of spallation $t_\mathrm{sp}$ is determined by measuring the velocity of spall layer using electric-contact gauge. On calculated graph of the pressure at the spall plane as a function of time (see figure~5a) with taking into account the value $t_\mathrm{sp}$, the spall strength is determined as $\sigma_\mathrm{sp}=-P(t_\mathrm{sp})$. The strain rate is determined from the graph of density as a function of time (see figure~5b) by derivation of $\rho(t)$ with respect to time: $\dot{V}/V_0\simeq -\dot\rho(t_\mathrm{sp})/\rho_0$, where $\rho_0=1/V_0$, $\dot{\rho}$ is the rate of density changing with time.

\begin{figure}[t]
\centering{\includegraphics[width=0.85\columnwidth]{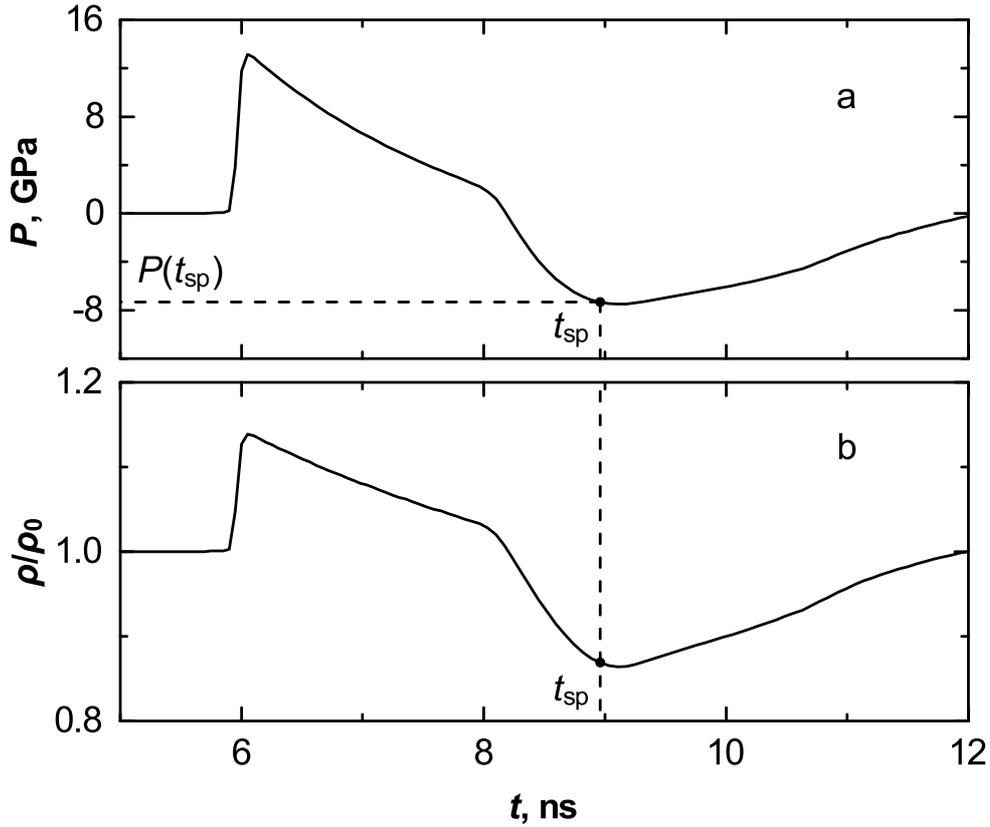}}
\caption{Calculated pressure (a) and normalized density of matter (b) at the spall plane in aluminum target of thickness 50~$\mu$m as functions of time. The laser pulse intensity is 3~TW/cm$^2$; the ablation pressure is 0.17~TPa; the pulse duration is 70~ps; $t_\mathrm{sp}$ denotes the moment of spallation; $\rho_0=2.71$~g/cm$^3$ is the normal density.}
\end{figure}

\subsection{Spall strength of aluminum targets}
Results of measurements of spall strength as a function of strain rate for aluminum and aluminum-magnesium alloy AMg6M~\cite{12a} are shown in figure~6. One can see that, at strain rate 56~$\mu$s$^{-1}$, dynamic strength of both materials achieves the ultimate value about 10.5~GPa according to the equation-of-state model~\cite{16b}. Theoretical estimations by method of full-potential linearized muffin-tin orbital~\cite{18} give the ultimate strength of aluminum 11.7~GPa. So, the experimental value of ultimate strength of aluminum is on 11\% less than theoretical evaluation~\cite{18}. In figure~6, previous data~\cite{19, 19a} for the two materials are also presented.

In experiments with aluminum targets, besides direct laser action, irradiation by the 2.5~ns pulses were used for acceleration of flyer plates (Al) of thickness 8 and 15~$\mu$m those impact the targets~\cite{Vovchenko-QE-2007, Vovchenko-PF-2009}. That allowed shortening of the shock-wave pulse down to 1.4~ns and advancing to higher strain rate domain.

\begin{figure}[t]
\centering{\includegraphics[width=0.85\columnwidth]{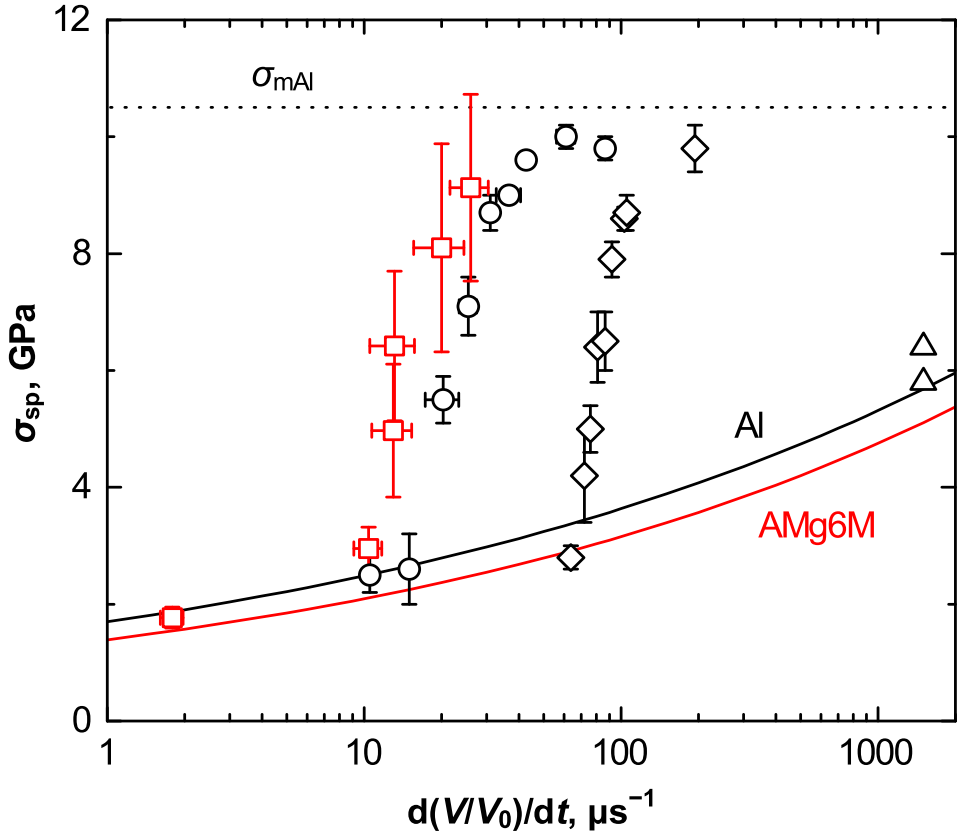}}
\caption{Spall strengths of the alloy AMg6M (squares, $\tau=2.5$~ns) and aluminum (circles, $\tau=2.5$~ns; diamonds, $\tau=70$~ps \cite{12a}; triangles, $\tau=150$~fs \cite{19a}) as functions of strain rate. Solid and dashed lines correspond to the functions for AMg6M and aluminum according to previous data~\cite{19}: $\sigma_\mathrm{sp}=1.39(\dot{V}/V_0)^{0.178}$ and $1.7(\dot{V}/V_0)^{0.165}$, respectively. Dotted line corresponds to ultimate spall strength $\sigma_\mathrm{mAl}=10.5$~GPa of aluminum according to the equation of state~\cite{16b}.}
\end{figure}

\subsection{Spallation of carbon modifications}
Results of investigations of the spall strength upon the strain rate for diamond~\cite{13, 13a, 14} and graphite~\cite{14a, 14b} are presented in figure~7. In experiments with synthetic diamonds, the value $\sigma_\mathrm{sp}\simeq 16.5$~GPa at the strain rate 70~$\mu$s$^{-1}$ is achieved, which is 24\% of theoretical evaluation of ultimate strength 69.5~GPa according to equation of state~\cite{17}. In the case of graphite, at the strain rate of 10~$\mu$s$^{-1}$, the spall strength $\sigma_\mathrm{sp}\simeq 2.1$~GPa is obtained that is 64\% of theoretical ultimate value 3.27~GPa from equation of state for the graphite~\cite{17}.

\begin{figure}[t]
\centering{\includegraphics[width=0.85\columnwidth]{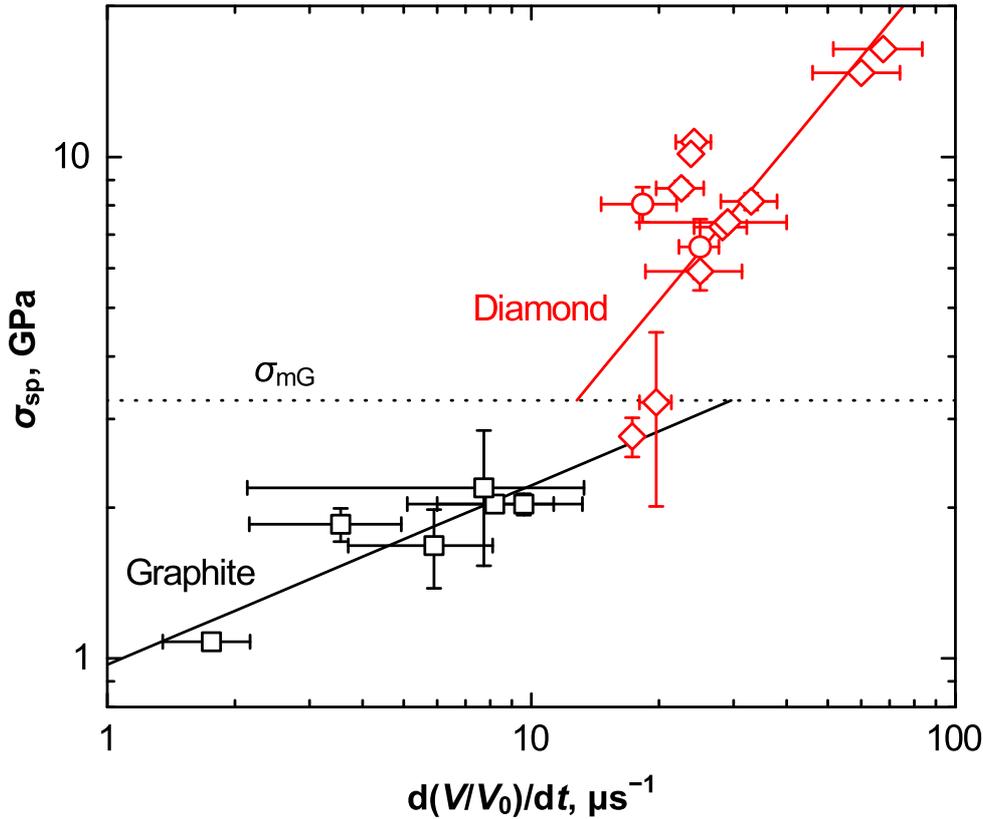}}
\caption{Spall strengths of diamond (circles---synthetic monocrystal; diamonds---synthetic polycrystals) and graphite (squares) as functions of strain rate. Solid and dashed lines correspond to the functions $\sigma_\mathrm{sp}=0.243(\dot{V}/V_0)^{1.02}$ for diamond and $\sigma_\mathrm{sp}=0.971(\dot{V}/V_0)^{0.358}$ graphite. Dotted line corresponds to ultimate spall strength $\sigma_\mathrm{mG}=3.27$~GPa of graphite according to the equation of state~\cite{17}.}
\end{figure}

\subsection{Polymorphic transformations of diamond and graphite}
Phase composition of carbon near spall region in targets of graphite~\cite{14a} and synthetic diamond~\cite{13a} was studied by method of Raman spectroscopy using spectrometer LabRam~HR (Horiba) with a spectral resolution 0.5~cm$^{-1}$. It was evidenced that, near the spall region of targets of polycrystalline diamond, some part of the material transforms to disordered graphite-like phase at deformation and spallation of diamond layer.

Raman spectra in the spall region of graphite targets indicate the well re-crystallized structure of the graphite with small disordering and higher degree of crystallinity than it was initially.

\subsection{Spall strength of polymethylmethacrylate targets}

Direct laser interaction and laser-driven thin foils (shock impact) were used for investigation spallation phenomena in polymethylmethacrylate (PMMA) targets in case of high strain rate~\cite{15}. The aluminum foils of thickness 8 and 15~$\mu$m were used for the impact. Mass and velocity of the laser-driven foils after laser ablation and acceleration were determined by the method of foil deceleration in a gas atmosphere. Basing on experimental data, we determined the spallation plane position (figure~8) and the time for the spall layer to arrive at an additional electric-contact gauge located beside the rear side of the target. Then the spall strength and strain rate have been calculated numerically using the hydrodynamic code with a wide-range semiempirical equation of state for PMMA~\cite{17}. The results of experiments are shown in figure~9.

\begin{figure}[t]
\centering{\includegraphics[width=0.85\columnwidth]{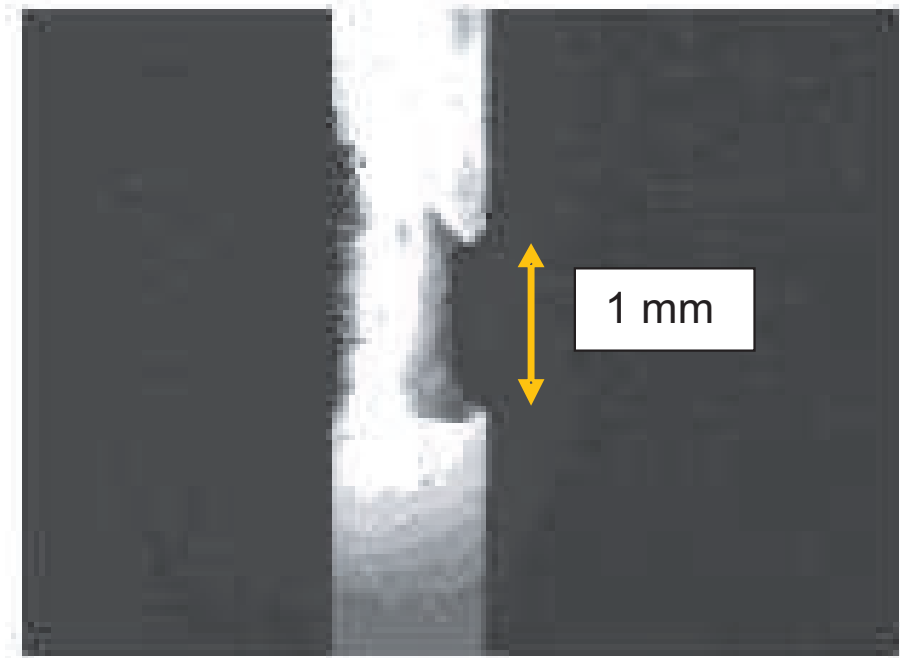}}
\caption{A target section after shock-wave action (face surface of the target is on the left-hand side).}
\end{figure}

\begin{figure}[t]
\centering{\includegraphics[width=0.85\columnwidth]{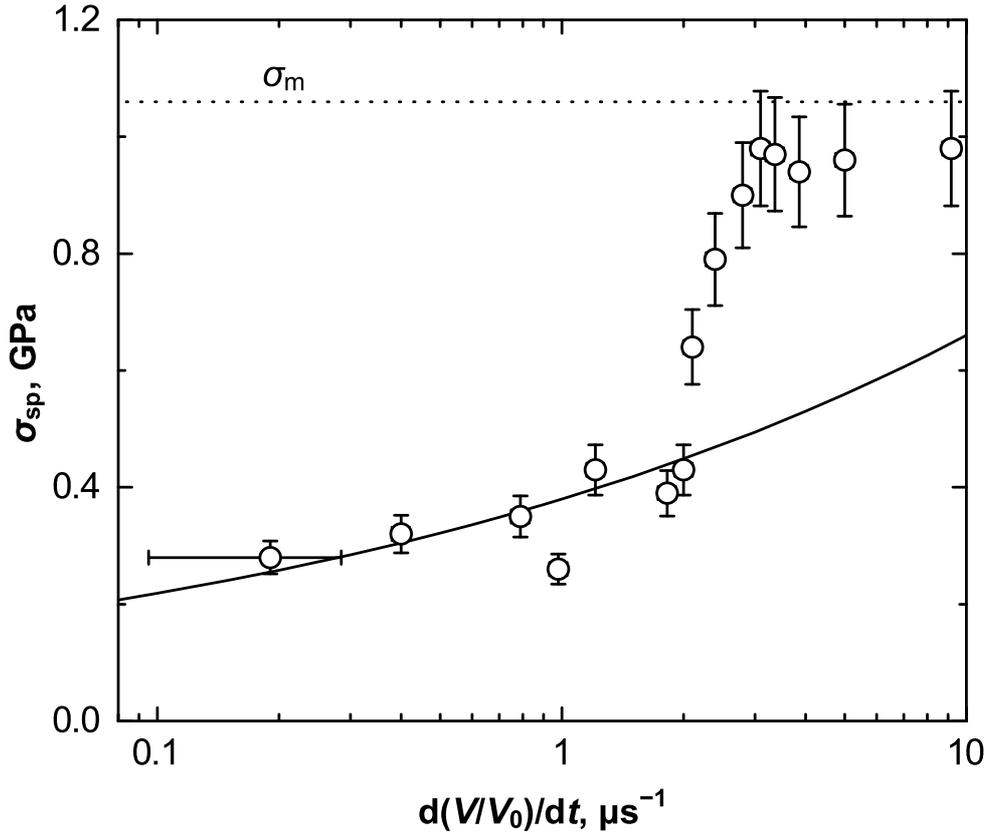}}
\caption{Spall strength of PMMA versus strain rate: circles correspond to data~\cite{15}; solid line represents a function of the spall strength of PMMA upon the strain rate $\sigma_\mathrm{sp}=0.38(\dot{V}/V_0)^{0.24}$ according to previous data~\cite{19}. Dotted line corresponds to ultimate spall strength $\sigma_\mathrm{m}=1.06$~GPa according to the equation of state~\cite{17} for PMMA.}
\end{figure}

It is shown that ultimate spall strength of PMMA has been reached. This ultimate strength value obtained in our experiments appears to be slightly below the theoretical estimation 1.06~GPa from the used equation of state.

As a result of the experiments, we have shown for the first time that the ultimate spall strength of PMMA (about 1~GPa) is achieved in case of strain rate 3~$\mu$s$^{-1}$.

\section{Conclusion}
Experimental studies at laser facilities of the A~M~Prokhorov General Physics Institute RAS have maiden essential contribution to high-energy-density physics and shown a possibility of modeling of astrophysical phenomena under laboratory conditions. The studies result in new data on thermophysical and mechanical properties of some materials in laser-generated shock waves and reveal prospects of achieving thermonuclear temperatures in conical targets~\cite{20}.

\ack
This work is performed under financial support by grants from the Russian Foundation for Basic Research (No.\,14-08-00967, 15-02-03498 and 16-02-00371) and the President of the Russian Federation (No.\,NSh-451.2014.2 and NSh-10174.2016.2), as well as the programs of the Presidium RAS (No.\,I.25P ``Fundamental and applied problems of photonics and physics of new optical materials'' and I.11P ``Thermophysics of high energy density'').

\end{document}